\documentclass[usenatbib]{basi}
\usepackage{amsmath,amsfonts,amssymb}
\usepackage{subfigure}

\usepackage[T1]{fontenc}
\usepackage[british]{babel}
\usepackage[varg]{txfonts}

\usepackage{rotating}
\usepackage{dcolumn}

\begin{document}

\title[detectability of binary pulsars]{The detectability of eccentric binary pulsars}

\author[M.~Bagchi et~al.]%
       {M.~Bagchi$^{1, \, 2}$\thanks{email: \texttt{manjari.bagchi@icts.res.in}},
       D.~R.~Lorimer$^{2}$ and S. Wolfe$^2$\\
       $^1$International Centre for Theoretical Sciences - Tata Institute of Fundamental Research, Bangalore 560012, India\\
       $^2$Department of Physics, White Hall, West Virginia University, Morgantown, WV 26506, USA}

\pubyear{2012}
\volume{00}
\pagerange{\pageref{firstpage}--\pageref{lastpage}}

\date{Received --- ; accepted ---}

\maketitle

\label{firstpage}

\begin{abstract}
Studies of binary pulsars provide insight into various theories of physics. Detection of such systems is challenging due to the Doppler modulation of the pulsed signal caused by the orbital motion of the pulsar. We investigated the loss of sensitivity in eccentric binary systems for different types of companions. This reduction of sensitivity should be considered in future population synthesis models for binary pulsars. This loss can be recovered partially by employing the `acceleration search' algorithm and even better by using the `acceleration-jerk search' algorithm.

\end{abstract}

\begin{keywords}
   stars: neutron, pulsars: general
\end{keywords}

\section{Introduction}\label{s:intro}
Binary pulsars are useful tools to get insight into various theories of physics, including theories of stellar evolution in binaries and theories of gravity. Detection of such systems in pulsar searches is challenging because of the Doppler modulation of the pulsed signal due to the orbital motion. In the case of Fourier domain searches for periodicities, the power in each harmonic spreads in adjacent bins. Thus, the height of the peak decreases, i.e. the observed flux degrades.  

\section{Formalism, results and future applications}\label{s:all}
\begin{figure}[h]
\vskip -0.4cm
 \begin{center}
\hskip -2cm \subfigure[$\gamma_1$ for NS-WD binaries]{\label{subfig:gamma1_nswd}\includegraphics[width=0.42\textwidth,angle=0]{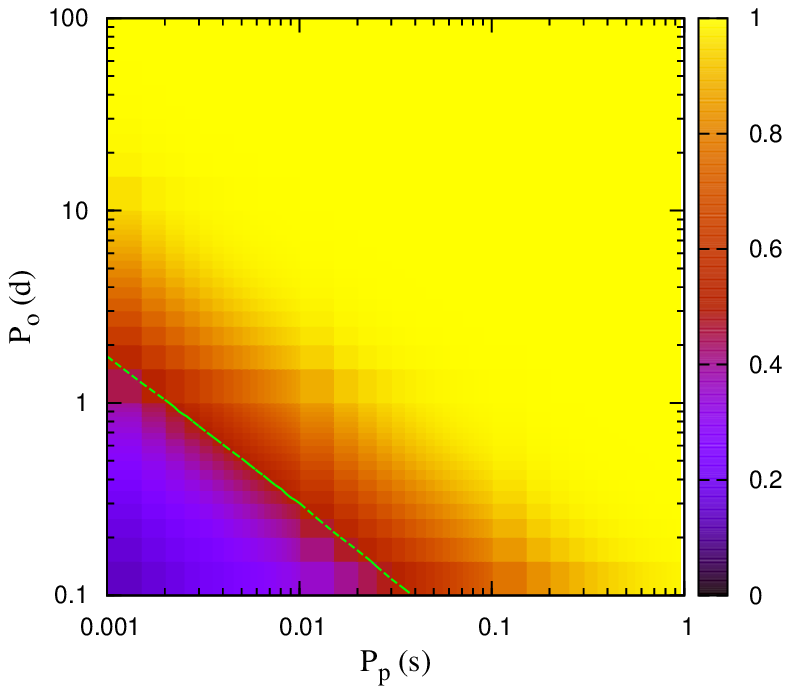}}
\hskip -1cm \subfigure[$\gamma_1$ for NS-NS binaries]{\label{subfig:gamma1_nsbh}\includegraphics[width=0.42\textwidth,angle=0]{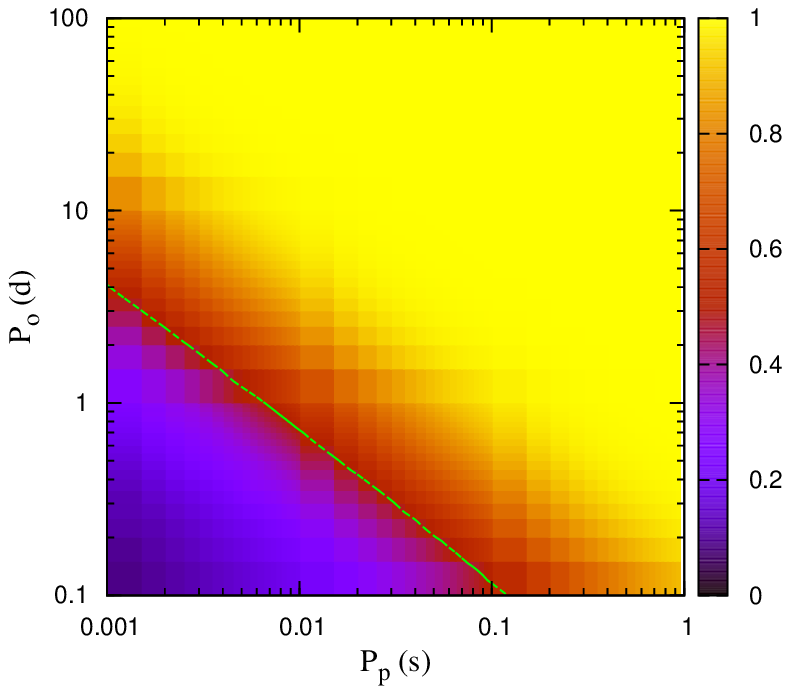}}
\vskip -0.5cm \hskip -2cm \subfigure[$\gamma_2$ for NS-WD binaries]{\label{subfig:gamma2_nswd}\includegraphics[width=0.42\textwidth,angle=0]{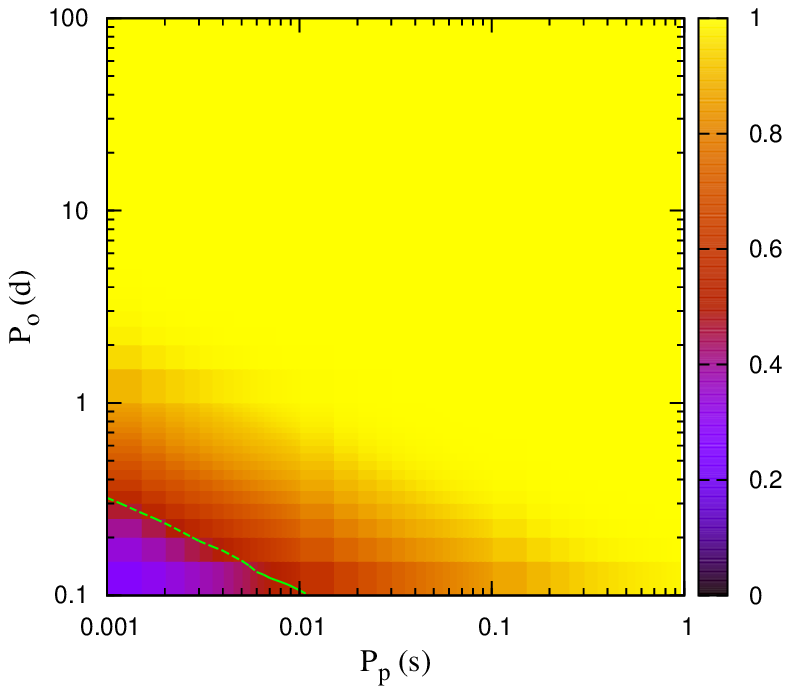}}
\hskip -1cm \subfigure[$\gamma_2$ for NS-NS binaries]{\label{subfig:gamma2_nsbh}\includegraphics[width=0.42\textwidth,angle=0]{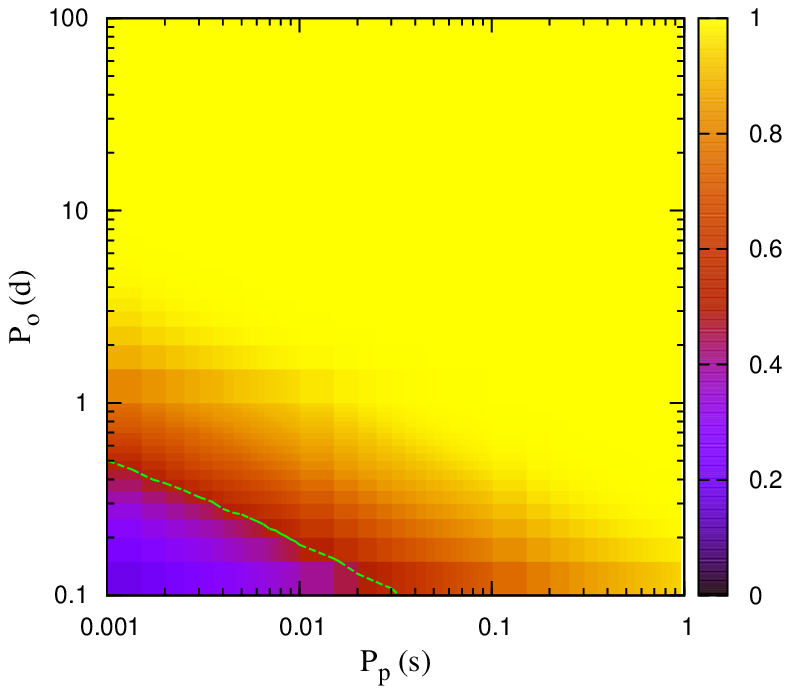}}
 \end{center}
 \vskip -0.5cm
\caption{Flux degradations without and with the acceleration search algorithm for NS-WD and NS-NS binaries. The X and Y axes represent the values of the spin and the orbital periods respectively. The color code represent the degradation factors with the contour for $0.5$ visible. The degradations are higher for NS-NS binaries than NS-WD binaries. For both the cases, $\gamma_2$ is greater than $\gamma_1$ provided other parameters are the same. The neutron star mass was chosen as $1.4~{\rm M_{\odot}}$, the white dwarf mass was chosen as $0.3~{\rm M_{\odot}}$.}
\label{fig:gammaall}
\end{figure}

The degradation of the flux due to the orbital motion of the pulsar can be quantified by a factor $\gamma$ such as $S_{obs} = \gamma^2 \, S_{int}$ where $S_{int}$ is the intrinsic flux of the pulsar and $S_{obs}$ is the observed flux. One can define three degradation factors, $\gamma_{1}$ without any intelligent search algorithm, $\gamma_{2}$ with the `acceleration search' algorithm, and $\gamma_{3}$ with the `acceleration-jerk search' algorithm. Smaller values of these parameters imply larger degradation. Johnston \& Kulkarni (1991) derived analytical expressions for $\gamma_{1}$ and $\gamma_{2}$ for pulsars in {\em circular orbits}. We computed these expressions, as well as the expression for $\gamma_{3}$ for {\em eccentric binaries}, as now there are a number of eccentric binaries - including the neutron star$-$neutron star binaries and the binaries in globular clusters. Analytical expressions for these factors, methods to compute, and explorations of the parameter space can be found in Bagchi, Lorimer, Wolfe (2013). The degradation increases for heavier companions and smaller spin and orbital periods, and decreases for higher eccentricities. For any particular set of parameters, $\gamma_3 > \gamma_2 > \gamma_1$. As an example, in Fig. \ref{fig:gammaall}, we show variations of $\gamma_1$ and $\gamma_2$ with the spin and the orbital periods for neutron star$-$white dwarf (NS-WD) and neutron star$-$neutron star (NS-NS) binaries. The degradation is larger for NS-NS binaries. So, we expect it to be even higher for neutron star$-$black hole binaries, which might be one of the reasons for non detection of any such binary till date, and the use of an `acceleration-jerk' search algorithm might be useful for this purpose. 

\appendix

\end{document}